# Micromechanical approach for the analysis of wave propagation in particulate composites

# Modelamiento micro-mecánico de la propagación de ondas en materiales compuestos con refuerzos esféricos

Cristhian Fernando Rojas-Cristancho[1a], Florence Dinzart[1b], Octavio Andrés González-Estrada[2]

[1]Laboratoire d'Etude des Microstructures et de Mécanique des Matériaux (LEM3), Université de Lorraine, 1 route d'Ars Laquenexy, 57078 Metz, France.
Email: [a] cristhian-fernando.rojas-cristancho1@etu.univ-lorraine.fr, [b] florence.dinzart@univ-lorraine.fr
[2] Grupo de Investigación en Energía y Medio Ambiente (GIEMA), School of Mechanical Engineering, Universidad Industrial de Santander, Colombia. Email: agonzale@saber.uis.edu.co

**Abstract**

Laser ultrasonic non-destructive testing is widely used for the inspection of mechanical structures. This method uses the propagation of ultrasonic guided waves (UGW) in the media. For this purpose, it has been demonstrated that the addition of a thin composite layer between the laser source and the structure for inspection is necessary. Consequently, this composite is an optoacoustic transducer composed of an absorption material such as carbon for inclusions and an expanding material such as an elastomer for the matrix. Thus, optimal fabrication of this composite should enable the amplification of the signal for inspection. Indeed, experimental research has demonstrated that variation in the volume fraction of carbon inclusions, their shape, and the nature of the matrix, affect the amplification of the signal directly. The aim of this study is to analyse the wave propagation in particulate viscoelastic composites by a dynamic self-consistent approach.

**Keywords:** particulate composites; self-consistent; viscoelastic composites; wave propagation.

**Resumen**

La inspección de componentes mecánicos por ultrasonido láser es uno de los controles no destructivos (CND) más utilizados en la industria, ya que permite inspeccionar rápidamente piezas de gran tamaño y de formas complejas por medio de la propagación de ondas guiadas. Ha sido demostrado que, para obtener la mejor calidad posible de la señal acústica, es necesario integrar una fina capa de material compuesto entre la placa y la fuente laser. Dicha capa de material compuesto permitiría la amplificación de la señal acústica, esta capa está formada por refuerzos de carbono que dan una característica de absorción térmica y de una matriz elastómera que otorga una característica de expansión volumétrica. Por lo tanto, la fabricación óptima de dicho compuesto permitiría la amplificación de la señal de inspección. De hecho, experimentalmente ha sido demostrado que la variación de la fracción volumétrica del refuerzo, de su forma (esférica o elipsoidal) y del tipo de matriz (silicona o resina), afecta directamente la amplificación de la señal. El objetivo de este trabajo es realizar un estudio micro-mecánico de tipo auto-coherente de la propagación de ondas elásticas en un medio heterogéneo compuesto por una matriz viscoelástica y refuerzos esféricos elásticos.

## 1. Introduction

Reliability in mechanical engineering structures after fabrication depends on how well these are controlled in a lifetime period [1], [2]. For that, non-destructive testing (NDT) is a useful tool. Ultrasonic guided waves (UGW) for inspection in mechanical structures have been widely investigated and developed [3]. This is because this method allows for evaluating large and complex form structures with only one measurement, unlike the conventional ultrasonic method using a transmitter-receiver system for local measurement.

A variation for UGW generation consists in using the power density of an incident laser beam [4]. For low power densities, ultrasonic waves are caused by the rapid thermal expansion of the material being irradiated. For this purpose, an optoacoustic transductor must transform in an easy and efficient way laser energy into acoustic (elastic) waves. This method to generate mechanical energy has been the focus of research in [5]–[9], where many different configurations of composite materials such as gold nanoparticles, carbon nanofiber and carbon nanotubes have been studied experimentally to obtain high-performance laser ultrasound transducers. In the same way, Ref. [10], [11] have focused on viscoelastic particulate composites.

According to [10], carbon nanoparticles produced by candle soot (CSPs) is an efficient light absorbing material and viscoelastic Polydimethylsiloxane (PDMS) has a high thermal coefficient of volume expansion. Therefore, CSPs – PDMS composite is found to be a high-performance optoacoustic transducer, since it could generate a high frequency, broadband and high amplitude ultrasound wave.

Inspired by the aforementioned results, we are interested in carrying out an analysis of elastic wave propagation in heterogeneous media. A first theoretical thermo-acoustic validation is proposed in [10], where the generated pressure gradient is directly associated with the temperature gradient generated by the laser impact. Under meaningful assumptions, the pressure gradient may be related solely to the setting parameters of the laser source.

Another analysis consists in considering the propagation process being in steady state. Then, it is possible to consider elastic behaviour and temperature gradient separately, as proposed in [12]. In this case, the displacement field of the propagating wave is supposed coincident with the displacement field of solid media. Therefore, the solution searched is the eigenvalues and eigenvectors velocities of the propagating wave in the solid media.

Solving the wave propagation problem is classic for homogeneous media [12]. The micromechanical approach is a traditional way to estimate the effective behaviour of heterogeneous materials, especially for static problems [13]. Within the framework of dynamic problems, acceleration term in the Navier equation has to be retained to evaluate the dynamic response of the composite through a micromechanical self-consistent approach [14].

The dynamic response in a solid homogeneous media for the wave propagation problem must be evaluated by the wave number $k_\gamma$, where $\gamma = \alpha$ for longitudinal waves and $\gamma = \beta$ for shear waves. This wave number belongs to the complex domain:

$$k_\gamma = \frac{\omega}{V(\omega)} + i\,\alpha(\omega) \qquad (1)$$

where, $V(\omega)$ and $\alpha(\omega)$ are the phase propagation velocity and the propagation attenuation coefficient, respectively. Dissipative properties of viscoelastic composites directly affect the attenuation coefficient as demonstrated experimentally in [15], [16].

This work presents in section 2 a general micromechanical self-consistent formulation based in [14], for analysing the elastic wave propagation in a heterogeneous solid media. Multiple scattering caused by the random distribution and interaction between particles complicates considerably the wave propagation analysis. As a consequence, we considered a single scattering problem described in section 2. Afterwards, the addition of viscoelastic properties to the micromechanical formulation is presented. In section 3, results obtained are presented and discussed. Finally, in section 4 some conclusions and limitations of the study are presented.

## 2. Methodology

### 2.1. Dynamic self-consistent formulation

Micromechanical methods are widespread to estimate composite behaviour within the static framework. For the dynamic case where acceleration terms are considered, the perturbation in the infinite media is supposed to be the mean propagating wave. Moreover, volume inclusion concentration in composites is an available information in practice, and not the distribution and interaction between particles. Accordingly, a first approach consists in considering the single inclusion problem embedded in the matrix possessing effective media properties.

A composite is considered comprising a matrix with elastic moduli tensor $L_{n+1}$ and density $\rho_{n+1}$, in which are embedded $n$ different types of inclusions, an inclusion of family phase $r$ has an elastic modulus tensor $L_r$ and density $\rho_r$. Volume concentration in the composite is such that:

$$\sum_{r=1}^{n+1} c_r = 1. \tag{2}$$

Solving the dynamic problem consist in determining a mean response to the displacement field on the solid $<u>$. For this purpose, the starting point is the Navier equation, that in the absence of body force reads:

$$div\ \sigma = \dot{p}, \tag{3}$$

where $\sigma$ is the stress field and $p$ is the momentum density. Stress and momentum density are related to elasticity Hooke's law and displacement wave field respectively:

$$\sigma = Le, \qquad p = \rho\dot{u}. \tag{4}$$

At the same time, deformation $e$ is related to displacement field through Cauchy's law in elasticity for small displacement assumption.

Analysis of the composite behaviour by a classical micromechanical approach comprising $n$ phases, suppose that an effective response of the composite behaviour is the result of the micromechanical behaviour of each constituent, thus:

$$div <\sigma> = <\dot{p}>. \tag{5}$$

Correspondingly, other quantities describing composite behaviour are averaged, such that, we obtain:

Effective deformation field:

$$<e> = \sum_{r=1}^{n+1} c_r e_r \tag{6}$$

Effective displacement field:

$$<\dot{u}> = \sum_{r=1}^{n+1} c_r \dot{u}_r \tag{7}$$

Effective stress field:
$$<\sigma> = \sum_{r=1}^{n+1} c_r \sigma_r = \sum_{r=1}^{n+1} c_r L_r e_r \tag{8}$$

Effective momentum density field:

$$<p> = \sum_{r=1}^{n+1} c_r p_r = \sum_{r=1}^{n+1} c_r \rho_r \dot{u}_r \tag{9}$$

Using perturbation theory, shown in [17] for the static problem, elastic and momentum density field in (5) can be separated by splitting into a homogenous part (comparison media) and a fluctuating part (perturbation). Consequently, (8) and (9) are written as follows:

$$<\sigma> = L_{n+1} <e> + \sum_{r=1}^{n+1} c_r (L_r - L_{n+1}) e_r, \tag{10}$$

$$<p> = \rho_{n+1} <\dot{u}> + \sum_{r=1}^{n+1} c_r (\rho_r - \rho_{n+1}) \dot{u}_r. \tag{11}$$

The final solution is then obtained for $e_r$ and $\dot{u}_r$ expressed as a function of the mean fields $<e>$ and $<\dot{u}>$. To this aim, the problem is simplified by introducing the single scattering problem for spherical inclusions, as presented thereafter.

## 2.2. Single scattering problem

This problem considers a single inclusion of volume $\Omega_r$ embedded in an elastic or viscoelastic effective matrix. The displacement field of the composite is coincident with the displacement field of the wave, that is represented through the wave propagation equation in solids:

$$<u> = m \exp[i(kx - \omega t)], \tag{12}$$

where $m$ is the wave amplitude, $k$ is the wave number and $\omega$ is the frequency. As a consequence, the dynamic response of an effective media of subscript '0', equation (13), must be equivalent to equations (10) and (11):

$$\sigma = L_0 e, \qquad p = \rho_0 \dot{u}. \tag{13}$$

Mathematical treatment that follows requires of perturbation theory, micromechanical Green's function, 'polarization' theory and operation of convolution. This development is well presented in previous works [14], [18]. Final equations represent the dynamic response of effective properties $(L_0, \rho_0)$ to the elastic wave propagation:

$$L_0 = L_{n+1} + \sum_{r=1}^{n} c_r h_r(k) h_r(-k) (L_r - L_{n+1}) \left[ I + \bar{S}_x^{(r)} (L_r - L_0) \right]^{-1} \tag{14}$$

$$\rho_0 = \rho_{n+1} + \sum_{r=1}^{n} c_r h_r(k) h_r(-k) (\rho_r - \rho_{n+1}) [I + \bar{M}_x^{(r)} (\rho_r - \rho_0)]^{-1} \tag{15}$$

where $h_r(k), h_r(-k)$ and $\bar{S}_x^{(r)}, \bar{M}_x^{(r)}$ are localization functions and dynamic tensors, respectively. They are dependent of the shape and size of the inclusion, the wave frequency, as well as effective properties. As shown in (14) and (15), the equation system is implicit and it must be solved by iteration.

In addition, the static response of a composite can be estimated from (14) and (15), considering the zero-frequency limit of equations, $\omega \approx 0$, $k = 0$, $h_r(k) = 1$ and $\bar{M}_t^{(r)} = 0$. Thus, equations are reduced to the classical micromechanical approach for the static problem:

$$L_0 = L_{n+1} + \sum_{r=1}^{n} c_r (L_r - L_{n+1}) A_r, \qquad (16)$$

$$\rho_0 = \sum_{r=1}^{n} c_r \rho_r, \qquad (17)$$

where $A_r$ are localization tensors in the classical micromechanical theory.

### 2.3. Spherical inclusions case

As proposed in section 2.2, equations (14) and (15) are dependent of the shape of the inclusion. In this section, it is presented the system of equations that describes the dynamic effective response to the elastic wave propagating in a spherical particle composite.

Since a heterogeneous media has become homogenised by a self-consistent approach, the effective media can be described with the aid of elastic moduli. Thus, the behaviour of the phase $r$ is represented by two elastic constants [12], $L_r = (3\kappa_r, 2\mu_r)$.

The dynamic response for a phase of random spherical inclusions, each of radius $a$, is as follows:

$$\kappa_0 = \kappa_2 + \frac{c_1 h_1(k) h_1(-k)(\kappa_1 - \kappa_2)}{1 + 3(\kappa_1 - \kappa_0)\varepsilon_\alpha/(3\kappa_0 + 4\mu_0)} \qquad (18)$$

$$\mu_0 = \mu_2 + \frac{c_1 h_1(k) h_1(-k)(\mu_1 - \mu_2)}{1 + \dfrac{2(\mu_1 - \mu_0)[2\mu_0\varepsilon_\alpha + (3\kappa_0 + 4\mu_0)\varepsilon_\beta]}{[5\mu_0(3\kappa_0 + 4\mu_0)]}} \qquad (19)$$

$$\rho_0 = \rho_2 + \frac{c_1 h_1(k) h_1(-k)(\rho_1 - \rho_2)}{1 + (\rho_1 - \rho_0)(3 - \varepsilon_\alpha - 2\varepsilon_\beta)/(3\rho_0)} \qquad (20)$$

Subscripts '1' and '2' represent the inclusion and matrix response. The equations (18-20) are function of volume inclusion concentration $c_1$, the functions $h_1(k)$ and $\varepsilon_\gamma$, which are presented for the spherical case below:

$$h_1(k) = 3(\sin ka - ka \cos ka)/(ka)^3 \qquad (21)$$

$$\varepsilon_\gamma = \frac{3(1 - i\,k_\gamma a)}{(k_\gamma a)^3}[\sin(k_\gamma a) \\ - k_\gamma a\,\cos(k_\gamma a)]e^{-ik_\gamma a} \qquad (22)$$

Subscript $\gamma$ denotes two possible polarizations for the wave propagation problem in homogeneous solid media, $\gamma = \alpha$ and $\gamma = \beta$ for the longitudinal and transversal wave propagation, respectively. Polarization for the wave propagation problem is shown in detail in [12] with the aid of the wave number in (1), as a consequence:

Longitudinal wave
$$k = k_\alpha = \omega[(3\kappa_0 + 4\mu_0)/3\rho_0]^{-1/2} \qquad (23)$$

Transverse wave
$$k = k_\beta = \omega(\mu_0/\rho_0)^{-1/2} \qquad (24)$$

Finally, (18–24) is a system of implicit equations, which is solved by iteration. First iteration is done by assigning to the effective properties the matrix properties ($L_0 = L_2, \rho_0 = \rho_2$). Because this study is interested in seeking the dynamic response of the composite adding viscoelastic properties, next section presents dynamic viscoelastic response ($L_2, \rho_2$) that will be integrated to the formulation in (18–20).

### 2.4. Viscoelastic properties

Viscoelastic homogeneous behaviour is represented in the complex domain, considering elastic-viscoelastic duality of material. Several rheological models have been proposed in the literature [19]. The chosen model to apply in the present study, for its simplicity, is the Maxwell rheological model. In this model, mechanical behaviour of the material is represented with the aid of a spring for elasticity behaviour and a damper for viscosity behaviour, both connected in series.

For homogeneous viscoelastic materials, the dynamic behaviour is described by two mechanical constants just as the homogeneous elastic material. This time, these properties are described in the complex domain. For this purpose, the fundamental equation stress-deformation for Maxwell materials,

$$\sigma_{ij}^* = \delta_{ij}\left[K^*(\omega) - \frac{2}{3}\mu^*(\omega)\right]\varepsilon_{kk}^* + 2\mu^*(\omega)\varepsilon_{ij}^* \qquad (25)$$

is introduced in (3). Since the solicitation remains in a propagating wave as (12), the equations obtained [12] for the dynamic response in viscoelastic materials are:

$$\lambda + 2\mu = \rho\left(\frac{2\pi f}{k_L}\right)^2 = \rho c_L^2 \left(1 + i\frac{c_L \alpha_L}{2\pi f}\right)^{-2} \qquad (26)$$

$$\mu = \rho \left(\frac{2\pi f}{k_T}\right)^2 = \rho c_T^2 \left(1 + i\frac{c_T \alpha_T}{2\pi f}\right)^{-2} \quad (27)$$

where, $c_T$ and $c_L$ denote the phase velocity, and $\alpha_T$ and $\alpha_T$ are the attenuation coefficients of the longitudinal and transverse waves, respectively.

### 2.4.1. Dynamic characterization of viscoelastic matrix

To calculate (26) and (27) to be introduced in the formulation model (18–20), viscoelastic properties of the matrix must be experimentally characterised through the dynamic values ($c_T$, $c_L$, $\alpha_T$ and $\alpha_L$) and are introduced as frequency dependent parameters defined by (26) and (27) in the propagation wave model (18-20).

Several previous experimental studies for the measurement of longitudinal wave propagation in viscoelastic materials have been carried out. It has been found that for a large frequency interval, phase velocity is invariant. Furthermore, the attenuation coefficient increases linearly with frequency ($\alpha = m\omega + \alpha_0$) [15]. This behaviour is presented in Table 1 and Figure 1. As EPOXY is often used as a matrix in the literature, we have also analysed the wave propagation problem with this material in section 3. EPOXY and PMMA experimental characterization are taken from [16]. In addition, an experimental characterization for the PDMS has been studied in [20].

Table 1. Dynamic characterization of viscoelastic matrix.

| Viscoelastic matrix material | Attenuation longitudinal coefficient $[np/cm]$ |
|---|---|
| PDMS 50:1 | $(3{,}538\,\omega[MHz] - 2{,}262) * (1/8{,}68589)$ |
| PDMS 10:1 | $(3{,}648\,\omega[MHz] - 1{,}36) * (1/8{,}68589)$ |
| EPOXY | $(45{,}4\,\omega[MHz] - 9{,}5) * (1/100)$ |
| PMMA | $(13{,}33\,\omega[MHz] - 6{,}67)$ |

## 3. Results

As mentioned before, the dynamic response to the wave propagation in a solid can be estimated with the wavenumber in (1). The phase velocity as well as attenuation coefficient are evaluated using (23,24).

Three special studies have been carried out in this work. First, the validation of micromechanical self-consistent, adding the viscoelastic properties to the formulation (18–20) developed in section 2, and presented in [14], where the formulation is solved for an elastic particulate composite. Second, once the model is validated, the model is subjected to the change of different parameters such as the volume concentration of inclusions and their size. Finally, these results are compared with experimental results, with [21] for phase velocity and [16] for the attenuation coefficient. The labels corresponding to the various configurations drawn in the following figures are presented in Table 2.

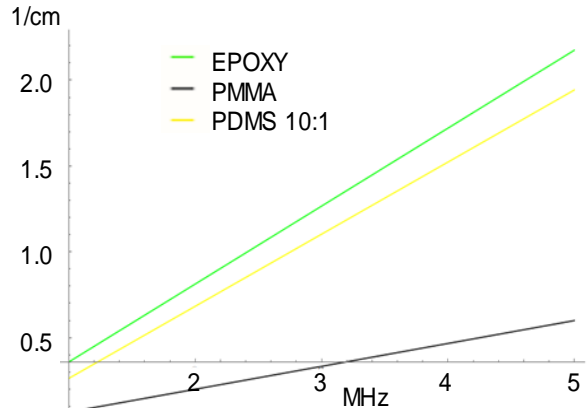

Figure 1. Dynamic characterization of viscoelastic matrix: EPOXY, PMMA, AND PDMS.

Table 2. Descriptions used in the figures.

| Label | Description |
|---|---|
| Viscoelastic matrix | Adding viscoelastic matrix response to (18–20) formulation. |
| Elastic Matrix | Elastic matrix response in (18-20). |
| Experimentation | Experimentation values [21]. |
| Theoretical Biwa work. | Work carried out by [16]. |

Table 3 summarizes the dynamic properties used for the materials in the study, taken from [14], [16]:

Table 3. Properties of the materials.

| Material | EPOXY | LEAD | GLASS |
|---|---|---|---|
| Longitudinal phase velocity (mm/μs) | 2,210 | 2,21 | 5,28 |
| Transverse phase velocity [mm/μs] | 1,197 | 0,86 | 3,24 |
| Density [g/cm3] | 1,202 | 11,3 | 2,47 |
| Shear modulus [GPa] | 1,731 | 8,357 | 26 |
| Bulk modulus [GPa] | 6,069 | 44,047 | 77.8 |

Figures hereafter have been normalised for phase velocity, attenuation coefficient and frequency. The phase velocity measurement is taken as $\omega/\mathrm{Re}\{k_\gamma\}$, where $\gamma = \alpha$ and $\gamma = \beta$ for longitudinal and transversal wave propagation , respectively. Where, $k_\gamma$ is defined by (23, 24). Phase velocities are normalised to the phase velocity

of a longitudinal wave propagation in the matrix material (C2L). The measure of attenuation is $\text{Im}\{a\,k_\nu\}$. Normalised frequency is $k_2 a$, where $k_2 = \text{Re}\{k_\alpha\}$ and is evaluated for $\kappa_0 = \kappa_2, \mu_0 = \mu_2, \rho_0 = \rho_2$.

### 3.1. Phase velocity

At first, the study focused on a composite with a large contrast in density between its constituents, for instance, an Epoxy-lead composite. It has been demonstrated that this large difference directly affects the dynamic response, mainly the effective density, which becomes to the complex domain [22].

### 3.1.1. Validation of the self-consistent model adding the dynamic viscoelastic response

Lead-epoxy composite is first considered, with size of inclusions set at $660\ [\mu m]$. Figures 2 and 3 show the study carried out for an epoxy matrix containing 5% and 15% of volume concentration spherical lead inclusions. The results are obtained for elastic and viscoelastic matrices. However, noticed that accounting for dynamic viscoelastic response leads to a global increase of the phase velocity. The experimental study carried out by Kinra [21] (yellow point in Figures 2 and 3) shows some peaks due to resonance phenomenon that are also clearly represented by the micromechanical approach. For this particular composite, resonance phenomenon is close to $k_\alpha a = 0.5$.

Notice that the micromechanical approach gives good approximations to experimentation, especially for upper-frequency values.

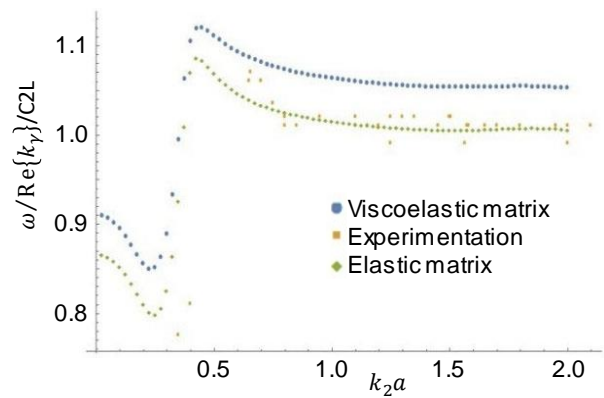

Figure 2. Phase velocity for longitudinal propagation, Epoxy – 5% Lead Composite.

Finally, the micromechanical self-consistent approach does not take into account the information on spatial correlations between inclusions [14]. Thus, this fact is a source of error and can be seen in the large difference in the resonance zone. In the literature, the self-consistent micromechanical approach gives a good approximation to low volume concentration of inclusions, typically 30% as in static problem.

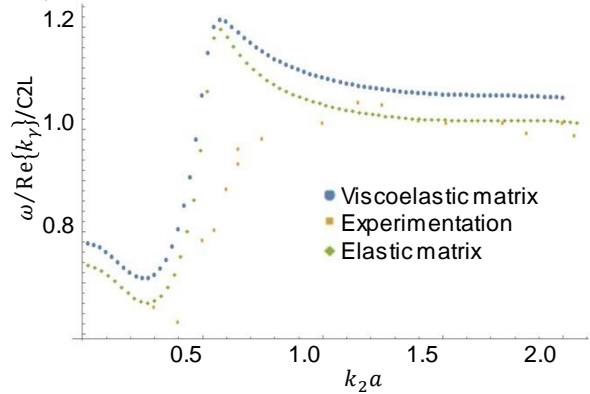

Figure 3. Phase velocity for longitudinal propagation, Epoxy – 15% Lead Composite

### 3.1.2. Variation of volume concentration inclusion

The experimental study in [21] for Epoxy – Lead particulate composite for longitudinal phase velocity shows a displacement of resonance phenomenon when volume concentration of inclusions increases (Figure 4).

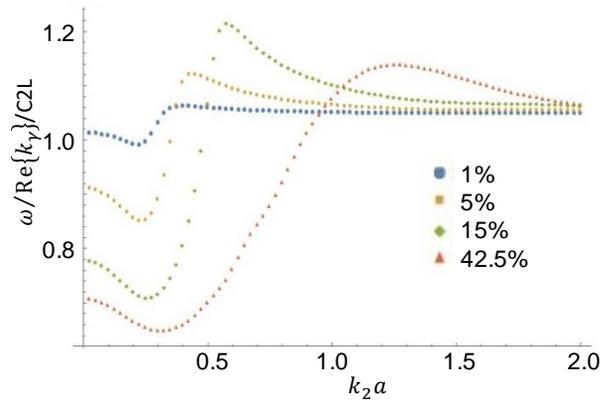

Figure 4. Longitudinal phase velocity for Epoxy - Lead composite. Variation of volume concentration.

### 3.2. Attenuation coefficient

In section 2.3, it has been shown a first characterization of attenuation coefficient in viscoelastic matrix (26) and (27) that has been necessary for the evaluation of dynamic effective properties in (18-20). In this section, imaginary part of (23, 24) are analysed.

### 3.2.1. Validation of self-consistent micromechanical model with viscoelastic matrix

Accounting for the viscoelastic behaviour of the matrix leads to change in the attenuation coefficient, in particular, at high-frequency regime. Attenuation coefficient becomes zero at high frequency for the elastic matrix. This attenuation coefficient has a defined value different from zero when viscoelasticity in the matrix is considered. Dynamic response of the effective media at high frequency corresponds to viscoelastic dynamic linear behaviour, as introduced in 2.4.1. The validation has been carried out for an Epoxy – Lead particulate composite, with the size of particles supposed at $a = 660\ [\mu m]$. Volume concentrations of 5 and 15% have been analysed in Figure 5.

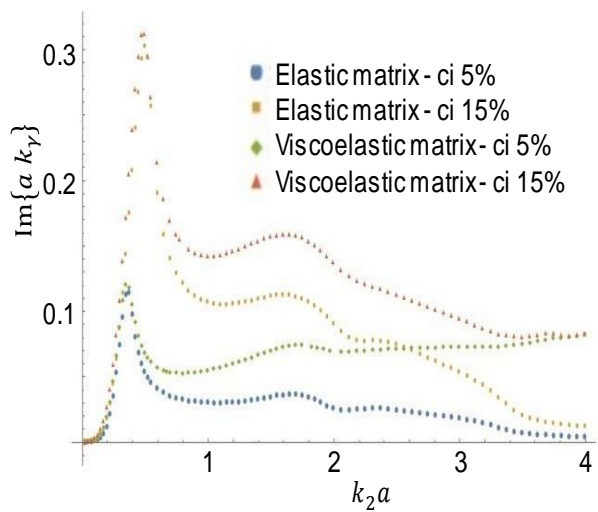

Figure 5. Longitudinal attenuation coefficient. Epoxy - 5% and 15% Lead.

The theoretical model developed in [16] evaluates attenuation coefficient by absorption due to viscoelastic behaviour and by scattering due to inclusions, separately. One of the particulate composites used for that study has been the Epoxy – Glass. In order to validate, this approach with the present work, Epoxy and Glass properties are taken from Table 3, and results are show in Figure 6.

Comparing Figures 5 and 6 it can be observed that the tendency of the curves is different for the same frequency range. Therefore, it is demonstrated that dynamic composite response depends on the inclusion nature.

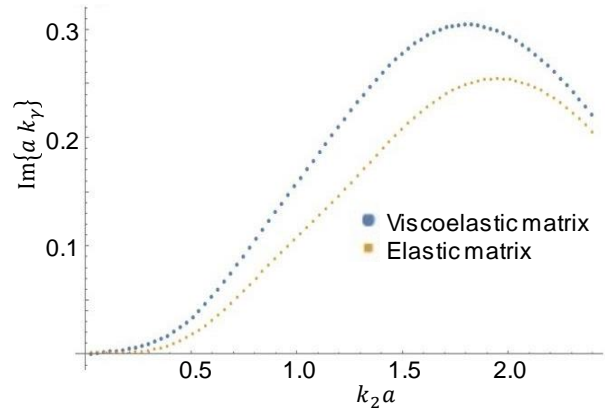

Figure 6. Longitudinal attenuation coefficient. Epoxy and 15% Glass.

### 3.2.2. Variation of volume concentration of inclusions

Figure 7 shows epoxy – lead composite with different values of volume concentration. At low volume concentration of inclusions, the attenuation coefficient seems to be linear due to viscoelastic behaviour. That means dissipation in wave propagation energy due to the absorption in the matrix. On the other hand, when the volume concentration increases, attenuation is due to the scattering in single inclusion.

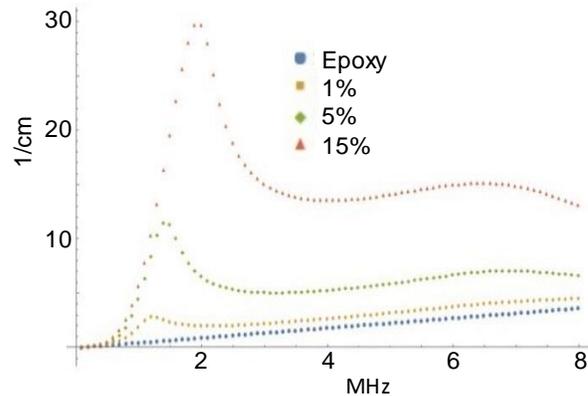

Figure 7. Longitudinal attenuation coefficient for Epoxy - Lead. Variation of volume concentration of inclusions.

### 3.2.3. Variation of the size of inclusions

As mentioned before, two factors cause attenuation of a wave propagating in a viscoelastic matrix composite. First, absorption due to the viscoelastic behaviour in the matrix and second, scattering due to the inclusions. In this section, the micromechanical dynamic self-consistent approach is compared to the theoretical model proposed by Biwa [16], who has studied the influence of size variation of inclusions in the attenuation coefficient.

Figure 8 shows the results obtained by Biwa for Epoxy – Glass composite, volume concentration of inclusions is 20%. It is worth to note that the attenuation coefficient in particulate composites is always higher than attenuation in pure viscoelastic material, contrary to the unidirectional fibre composites case [16]. Figure 9 shows the results for the micromechanical self-consistent approach.

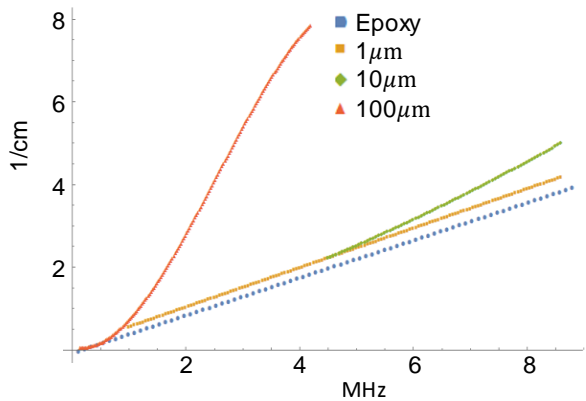

Figure 8. Longitudinal attenuation coefficient, theoretical approach Biwa [16]. Variation of the size of inclusions.

The tendency in Figures 8, 9 is similar when the size of the inclusions increases.

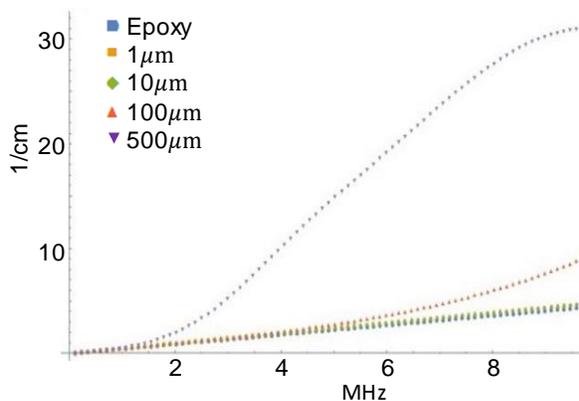

Figure 9. Longitudinal attenuation coefficient in micromechanical self-consistent approach. Variation of the size of inclusions.

The increase of the attenuation value is greater for larger size of inclusions and at low frequencies (Figure 8). However, for the dynamic – self-consistent approach

(Figure 9) this behaviour is rather observed at higher frequencies.

### 3.2.4. Experimental validation

The self-consistent model is compared to experimental results carried out by Kinra [21]. Attenuation coefficient for Epoxy - Glass composite has been analysed. The size of particles is 150 $[\mu m]$ and the volume concentration of inclusions is supposed to be 8,6%.

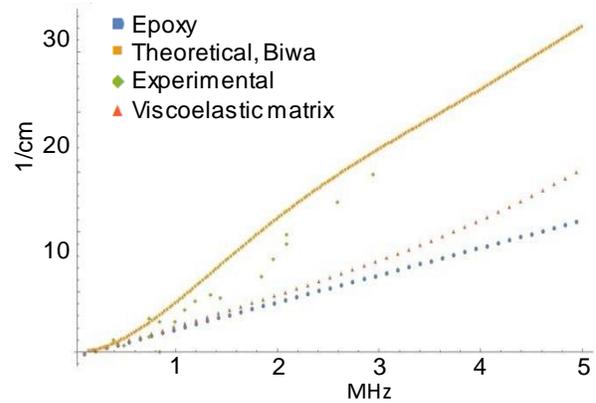

Figure 10. Experimental validation of attenuation coefficient. Epoxy - 8,6% Glass composite.

Both theoretical approaches presented in Figure 10 show a good approximation with experimentation work for a low volume concentration in inclusions (8,6%) and at low frequency. When these values increase, both approaches give an incorrect approximation, in the case of dynamic self-consistent, this is due to the no experimental information relating to the correlation between particles. Thus, the single scattering problem does not give a good approximation at high volume concentration values.

## 4. Conclusions

In this work, elastic wave propagation in a heterogeneous viscoelastic particulate media has been studied.

This approach is based on a dynamic micromechanical self-consistent approach. The dynamic term has been included in the Navier equation, it represents the displacement field of the wave which is coincident with the displacement field of effective media. In the literature, this approach gives good approximations at low concentration volume of inclusion, 30% maximum.

The dynamic micromechanical self-consistent approach does not take into account correlations between inclusions. Therefore, this is a source of error compared with experimental works. The dynamic composite response also depends on the inclusion nature.

The integration of dynamic viscoelastic response of matrix gives an increase in both, phase velocity and attenuation coefficient values, in comparison with the elastic model.